\documentclass[final]{jfm}

\usepackage{graphicx}
\usepackage{newtxtext}
\usepackage{newtxmath}
\usepackage{natbib}
\usepackage{hyperref}

\hypersetup{
    colorlinks = true,
    urlcolor   = blue,
    citecolor  = black,
}

\newcommand{\RomanNumeralCaps}[1]
\linenumbers

\title{Flow states and heat transport in liquid metal convection} 

\author{Lei Ren\aff{1},
  Xin Tao\aff{1},
   Lu Zhang\aff{3},
  Ming-Jiu Ni\aff{1,2},
  Ke-Qing Xia\aff{3},
 \and Yi-Chao Xie\aff{1}
 \corresp{\email{yichao.xie@xjtu.edu.cn}}}

\affiliation{\aff{1}State Key Laboratory for Strength and Vibration of Mechanical Structures and School of Aerospace, Xi'an Jiaotong University, Xi'an 710049, PR China
\aff{2}School of Engineering, University of Chinese Academy of Sciences, Beijing 100049, PR China 
\aff{3}Center for Complex Flows and Soft Matter Research and Department of Mechanics and Aerospace Engineering, Southern University of Science and Technology, Shenzhen 518055, PR China}

\begin{document}
\maketitle

\begin{abstract}
We present an experimental study of Rayleigh-B\'enard convection using liquid metal alloy gallium-indium-tin as the working fluid with a Prandtl number of $Pr=0.029$. The flow state and the heat transport were measured in a Rayleigh number range of $1.2\times10^{4} \le Ra \le 1.3\times10^{7}$. The temperature fluctuation at the cell centre is used as a proxy for the flow state. It is found that, as $Ra$ increases from the lower end of the parameter range,  the flow evolves from a convection state to an oscillation state, a chaotic state, and finally a turbulent state for $Ra>10^5$. The study suggests that the large-scale circulation in the turbulent state is a residual of the cell structures near the onset of convection, which is in contrast with the case of $Pr\sim1$, where the cell structure is replaced by high-order flow modes transiently before the emergence of the large-scale circulation in the turbulent state. The evolution of the flow state is also reflected by the heat transport characterised by the Nusselt number $Nu$ and the probability density function (PDF) of the temperature fluctuation at the cell centre. It is found that the effective local heat transport scaling exponent $\gamma$, i.e., $Nu\sim Ra^{\gamma}$, changes continuously from $\gamma=0.49$ at $Ra\sim 10^4$ to $\gamma=0.25$ for $Ra>10^6$. Meanwhile, the PDF at the cell centre gradually evolves from a Gaussian-like shape before the transition to turbulence to an exponential-like shape in the turbulent state. For $Ra>10^6$, the flow shows self-similar behaviour, which is revealed by the universal shape of the PDF of the temperature fluctuation at the cell centre and a $Nu=0.19Ra^{0.25}$ scaling for the heat transport.
\end{abstract}

\section{Introduction}\label{sec:int}
Turbulent convection in liquid metal occurs commonly in nature and in industry. Examples include the convection in the outer core of the Earth \citep{King2013} and the metallothermic titanium reduction reactor \citep{Teimurazov2017}. The kinematic viscosity $\nu$ of liquid metal is much smaller than its thermal diffusivity $\kappa$, resulting in the Prandtl number $Pr=\nu/\kappa$ being much smaller than 1. An idealised system to study thermal convection is the classical Rayleigh-B\'enard convection (RBC) system, i.e., a fluid layer confined between two plates heated from below and cooled from above. The flow state in RBC is controlled by the Rayleigh number $Ra=\alpha g\Delta T H^3/(\nu\kappa)$ and $Pr$. The aspect ratio $\Gamma$ serves as the third control parameter, quantifying the effect of geometrical confinement. For the most widely studied cell geometry, i.e., cylindrical cells, the aspect ratio is defined as $\Gamma = D/H$. Here, $\alpha$, $g$, and $\Delta T$ are the thermal expansion coefficient, the gravitational acceleration constant and the temperature difference across the cell; $H$ and $D$ are, respectively, its height and diameter. The heat transport of the system is quantified by the Nusselt number $Nu=qH/(\lambda \Delta T)$, which is the ratio of the heat flux due to convection to that would be transported solely by conduction. Here $q$ is the input heat flux supplied at the bottom plate and $\lambda$ is the thermal conductivity of the working fluid.

Recent studies on liquid metal convection mainly focus on the dynamics of the large-scale circulation (LSC) in the turbulent state. Combining experimental and numerical studies, \cite{Vogt2018} reported a new "jump-rope-vortex" mode of the LSC for $\Gamma=2$ cylindrical cell. This new flow mode was reported recently by \cite{Eckert2022} in square cross-section cells. By comparing the structure of the LSC in cylindrical cells with $\Gamma=1$ and 0.5, it is demonstrated that the single-role LSC observed in a $\Gamma=1$ cell will collapse in a $\Gamma=0.5$ cell, resulting in dependence of the $Nu\sim Ra$ scaling on $\Gamma$ \citep{Schindler2022}. The effects of cell tilting on the dynamics of the LSC and $Nu$ were also studied in liquid sodium convection \citep{Khalilov2018}. The above new phenomena add to our understanding of the dynamics of the large-scale flow (LSF) in turbulent convection. However, a fundamental issue remains unexplored, i.e., what is the origin of the LSC in liquid metal turbulent convection?  

It is well-known that the LSC is a LSF that emerges in the fully-developed turbulent state of RBC. When the flow will become turbulent also deserves study. Using gaseous helium ($Pr \sim 0.8$) as the working fluid, \cite{Heslot1987} demonstrated experimentally that the RBC system evolves from a conduction state to an oscillation state, a chaotic state, a transitional state, and a turbulent state with increasing $Ra$. The dynamics of the LSF in these states, i.e., the cell structure near the onset of convection and the LSC in the turbulent state, were studied by \cite{Wei2021} using compressed nitrogen gas. It is found that the strength of the cell structure increases with $Ra$, but the strength of the LSC decreases with $Ra$ in the turbulent state. Using direct numerical simulation (DNS), \cite{Verzicco1997} studied the flow state evolution in mercury, and showed that the flow is turbulent when $Ra \ge 3.75\times10^4$. However, the effects of $Pr$, especially in the low-$Pr$-regime, on the observed flow state evolution and the transitional Rayleigh number to turbulence have not yet been explored experimentally.

There has been evidence that the heat transport in liquid metal turbulent RBC is influenced by the LSF structure \citep{Zwirner2020, Schindler2022}. However, how the flow state evolution mentioned above affects heat transport remains unexplored. Early studies on heat transport in liquid metal convection were mainly performed in cells with $\Gamma>1$ \citep{Globe1959, Rossby1969, Horanyi1999, Aurnou2001JFM}. In cells with $\Gamma=1$, measurements using mercury with $Pr\sim0.025$ \citep{Takeshita1996,Cioni1997}, using gallium or gallium-indium-tin (GaInSn) alloy with $Pr\sim0.029$ \citep{King2013,Zurner2019},  and in DNS studies by \cite{Scheel2016} and \cite{Pandey2021}, all exhibit a $Nu\sim Ra^{0.25}$ scaling to the lowest order. In cells with $\Gamma<1$, i.e., elongated cylinders, studies in sodium convection with $Pr\sim 0.009$ found that the heat transport scaling exponents are larger than $0.25$ \citep{Frick2015, Mamykin2015}. However, a connection between the flow state evolution and the heat transport beyond the onset of convection is still missing, which motivates the present study. It is worth noting that different sidewall materials were used in the above experimental studies, i.e., stainless steel with $\lambda\sim$ 16 W/(mK) \citep{Takeshita1996, Cioni1997,Frick2015, Mamykin2015} and polyether ether ketone (PEEK) with $\lambda\sim 0.25$ W/(mK) \citep{Zurner2019}, resulting in the pre-factor of the heat transport scaling varies from 0.17 to 0.22. The coupling between the fluid and sidewall could affect the heat transport up to $20\%$ \citep{Ahlers2000}. Thus, measurements of the heat transport under more carefully designed sidewall boundary conditions (BCs) in liquid metal are necessary. To achieve this, a combination of Plexiglas ($\lambda\sim 0.192$ W/(mK)) with GaInSn ($\lambda\sim$ 24.9 W/(mK)) was chosen to mimic an adiabatic sidewall temperature BC in the present study.
 
In this paper, by measuring the flow states and the heat transport simultaneously, we will unravel how the flow state evolution influences heat transport in liquid metal convection. It will be shown that with increasing $Ra$, the flow in liquid metal convection evolves from convection to oscillation, chaotic, and turbulent states. In addition, the LSC in the turbulent state ($Ra>10^5$) is found to be a residual of the cell structure observed near the onset of convection. The flow evolution is also reflected by the heat transport of the system, i.e., the effective heat transport scaling exponent $\gamma$ changes continuously from $\gamma=0.49$ at $Ra\sim 10^4$ to $\gamma=0.25$ for $Ra>10^6$.

\section{The experimental setup and measurement techniques}\label{sec:exp}

The experiments were carried out in two $\Gamma=1$ cylindrical cells with heights of $H=$ 40 mm and 104 mm, respectively. They are labelled as cell A and cell B. The construction of the cells is similar to those used in \cite{Xie2017JFM}. Thus only their essential features will be described. Each cell consists of three parts, i.e., a heating bottom plate, a cooling top plate and a Plexiglas sidewall. The plates are made of oxygen-free copper. The sidewall made of Plexiglas with a thickness of 2.5mm (cell A) and 5mm (cell B) is sandwiched between the two plates. A fluorine rubber O-ring was placed in a groove carved into the top (bottom) surface of the bottom (top) plate to prevent the leakage of GaInSn. The bottom plate was heated by a nichrome wire heater. The top plate was cooled down by passing temperature-controlled water through double-spiral channels carved onto its top surface. Under such an arrangement, the temperature BC at the bottom plate is constant heat flux and that of the top plate is constant temperature, nominally. The respective temperatures of the top plate $T_t$ and the bottom plate $T_b$ were measured using four thermistors inserted into blind holes drilled from their side. These holes are distributed uniformly azimuthally along a circle with a radius of $D/4$. Another thermistor located at the centre of the bottom plate was installed.

GaInSn alloy was used as the working fluid with the mean fluid temperature $T_m=(\langle T_t\rangle+\langle T_b\rangle)/2$ kept at 35 $^o$C. Here $\langle \cdots \rangle$ represents time averaging. The physical properties of GaInSn at 35 $^o$C are: density $\rho=6.34\times 10^{3}$ kg/m$^{3}$, kinematic viscosity $\nu=3.15\times 10^{-7}$ m$^2$$/$s,  thermal conductivity $\lambda=24.9$ W$/$(mK), and thermal expansion coefficient $\alpha=1.24\times 10^{-4}$ $/$K \citep{Plevachuk2014}. Thus $Pr$ was fixed at $Pr=0.029$. By varying the temperature difference $\Delta T=\langle T_b\rangle-\langle T_t \rangle$ from 0.51 K to 26.91 K for cell A and from 1.13 K to 31.74 K for cell B, a $Ra$ range from $1.17\times10^{4}$ to $1.27\times10^{7}$ was achieved. The wall admittance is given by $C=[(D/2)\lambda]/(d_w\lambda_w)$ with $d_w$ being the thickness of the sidewall and $\lambda_w$ being the thermal conductivity of the sidewall \citep{Buell1983}. The present experimental setup gives $C=1038$ for cell A and $2248$ for cell B. Both values of $C$ are much larger than 1, suggesting that the sidewall temperature BC can be considered as adiabatic to a good approximation. Special care was taken when filling GaInSn into the convection cell to avoid oxidation of GaInSn. Argon gas was injected into the cell for 24 hours to remove oxygen. The fluid was then injected into the cell under the action of gravitation. The convection cell was wrapped by rubber foam with a thickness of 5 cm to prevent heat loss and was placed inside a thermostat to keep it isolated from the surrounding temperature variation.

A multi-thermal probe method was used to measure the dynamics of the LSF \citep{Cioni1997}. Thermistors located at distances $z=H/4$, $H/2$, and $3H/4$ from the bottom plate, with eight thermistors distributed uniformly azimuthally at each $z$, are used to measure the azimuthal temperature profile. At each $z$ and time step $t$, the temperature profile was decomposed into discrete Fourier modes according to 
\begin{equation}
T(i,t)= T_0(t) + \sum_{n=1}^{4} [a_{n}(t)\cos (\frac{n\pi}{4}i) + b_{n}(t)\sin (\frac{n\pi}{4}i)], (i=0 \dots 7)
\end{equation}
where $T(i,t)$ is the temperature of the $i^{th}$ thermistor at time instance $t$; $T_0(t)$ is the mean temperature of all thermistors at time instance $t$; $n=1, 2, ..., 4$ are the $n^{th}$ Fourier mode; $a_{n}$ and $b_{n}$ are the Fourier coefficients, from which the strength $\delta_n$ of the LSF can be determined using $\delta_n = \sqrt{{a_n}^2+{b_n}^2}$. The energy of the $n^{th}$ Fourier mode $E_n$ and the total energy of all modes $E_{t}$ are defined as $E_n={\delta_n}^2={a_n}^2+{b_n}^2$ and $E_{t}=\sum_{n=1}^{4} E_n$, respectively. If the LSF is in a single-roll form, $E_1/E_{t}$ should be close to 1. It has been shown by \cite{Zurner2019} that the structure and dynamics of the LSF measured directly using ultrasonic doppler velocimetry and those measured indirectly using the multi-thermal probe method are consistent with each other, suggesting that the multi-thermal probe method is capable of measuring the LSF in liquid metal convection.

A digital multimeter was used to measure the resistances of the 39 thermistors at a sampling rate of 0.34 Hz. The resistances were converted to temperatures using the parameters obtained from a separate calibration process. By measuring the voltage $V$ and current $I$ applied to the heater, the applied heat flux $q=4UI/ \pi D^2$ was obtained. For each $Ra$, $q$ was kept at a constant, $\Delta T$ and the dynamics of the LSF were measured simultaneously, from which we calculated $Ra$, $Pr$, $Nu$ and the flow modes. Another thermistor with a head diameter of 300 $\rm{\mu m}$ and a time constant of 30 $\rm{ms}$ was placed at the cell centre to measure the temperature in the fluid. This thermistor was sampled at a sampling rate of 20 Hz. Each measurement lasted at least 8 hours after the system had reached a steady state.

\section {Results and discussions}\label{sec:res}

\subsection {Flow states as a function of the Rayleigh number}\label{subsec1}

\begin{figure}
  \centerline{\includegraphics[width=\textwidth]{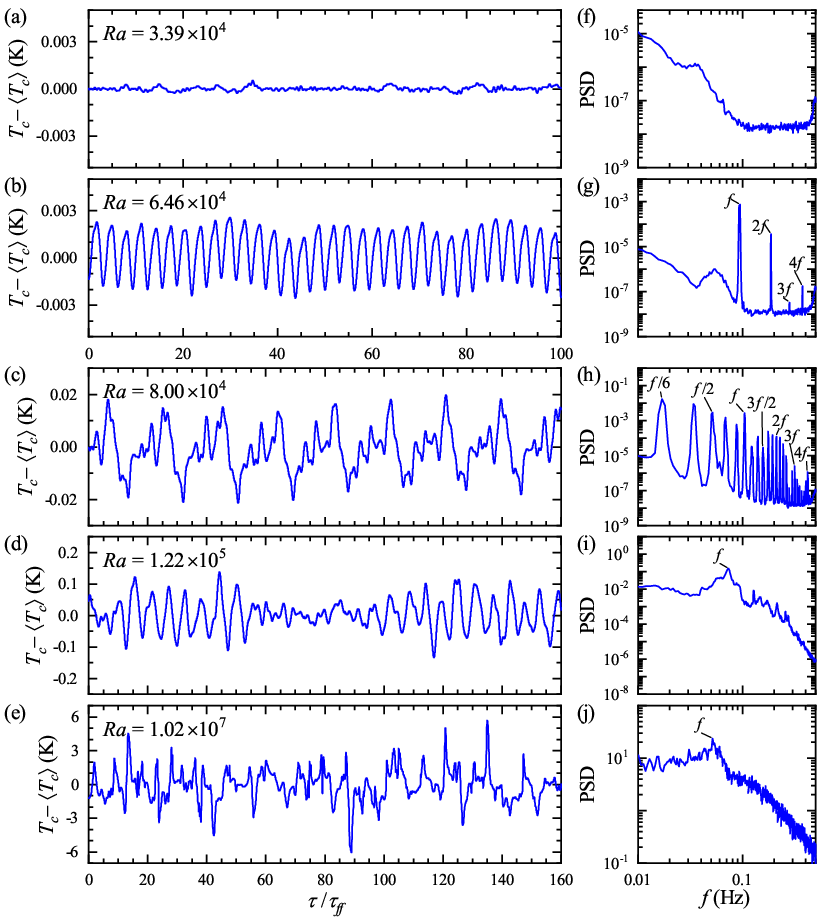}}
   \caption{Left panel: the time series of the temperature fluctuation $T_c-\left \langle T_c \right \rangle$ at the cell centre for different values of $Ra$. Right panel: the power spectral density (PSD) of the corresponding time series in the left panel.}
\label{fig1}
\end{figure}

We first discuss the evolution of the flow with increasing $Ra$. The temperature at the cell centre $T_c$ was used as a proxy of the flow state. Figure \ref{fig1}(a-e) show the time series of $T_c- \langle T_c \rangle$ for five values of $Ra$. The time axes are normalised by the free-fall time $\tau_{ff}=\sqrt{ H/\alpha g \Delta T}$. The corresponding power spectral density (PSD) of the time series is shown in figure \ref{fig1}(f-j). For $Ra=3.39\times10^4$, the $T_c$ is almost a constant (figure \ref{fig1}(a)), implying that the flow is stable and it is in the convection regime. A periodic oscillation of $T_c$ is observed for $Ra=6.46\times10^4$(figure \ref{fig1}(b)). This oscillation and its harmonics are also evidenced by the sharp peaks of the PSD (figure \ref{fig1}(g)). When $Ra$ increases to $Ra=8.00\times10^4$, a chaotic state is observed. From the PSD shown in figure \ref{fig1}(h), one sees not only the fundamental oscillation frequency $f$ and its harmonics but a new oscillation mode with $f/2$ and its harmonics. The results suggest that a period-doubling bifurcation occurs at this value of $Ra$ and the flow becomes chaotic. At $Ra=1.22\times10^5$ (figure \ref{fig1}(d)), the time series shows a combination of random fluctuations and regular low-frequency chaotic oscillations. The corresponding PSD shows that high-frequency components becoming more pronounced, signifying a turbulence signal. When $Ra$ increases to $1.02\times10^7$, sharp spikes appear in the time trace of $T_c$. The corresponding PSD becomes noisy in the full range of frequencies, similar to the PSD of temperature in turbulent RBC with working fluids like water \citep{Zhou2002PRL}, suggesting that the flow is now in a fully developed turbulent state.

\begin{figure}
  \centerline{\includegraphics[width=\textwidth]{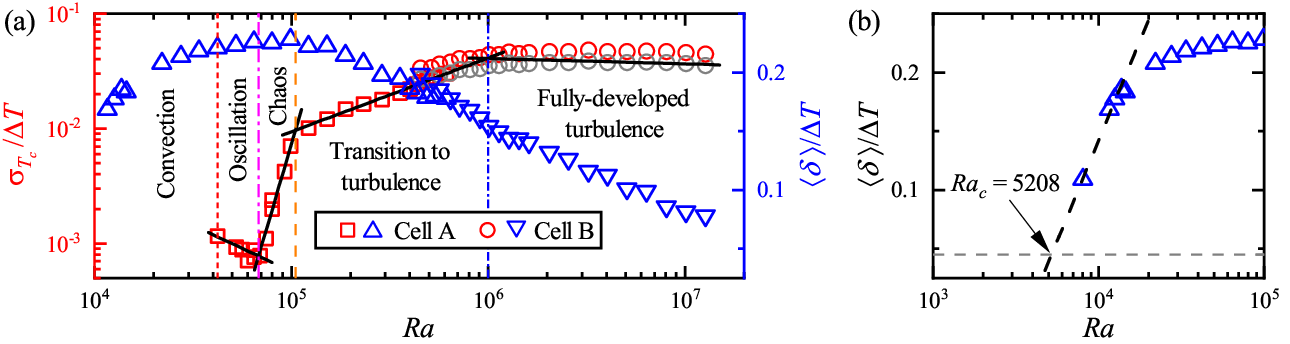}}
  \caption{(a) The normalised RMS temperature $\sigma_{Tc}$ at the cell centre vs $Ra$. Grey circles are multiplied by a factor of 0.8 to collapse measurements in cells A and B. The triangles are the flow strength of the first Fourier mode $\delta$ vs $Ra$. (b) $\delta$ vs $Ra$ near the onset of convection. The horizontal dashed line is the smallest $\delta$ measured when setting $\Delta T\sim 0$ K. The thick dashed line is a fit of $\langle\delta\rangle/\Delta T=0.15\ln Ra-1.22$, yielding $Ra_c=5208$. }
\label{fig2}
\end{figure}

The examples in figure \ref{fig1} demonstrate that the flow evolves from convection to an oscillation state, a chaotic state, and a turbulent state. The different states are characterised by different levels of temperature fluctuations. Next, we determine the transitional $Ra$ for different flow states based on the root-mean-square (RMS) temperature $\sigma_{T_c}=\sqrt{\langle(T_c-\langle T_c\rangle)^2\rangle}$. Figure \ref{fig2}(a) shows $\sigma_{T_c}/\Delta T$ vs $Ra$. For $Ra<4.2\times10^4$, there are hardly any temperature fluctuations. Thus the flow is in a steady convection state, as evidenced by the non-zero value of the amplitude of the first Fourier mode $\delta$ (blue triangles in figure \ref{fig2}). Several discrete transitions of $\sigma_{T_c}/\Delta T$ are observed in figure \ref{fig2}, implying that the flow experiences different states. To determine the transitional $Ra$ of the flow states, the data in different states are fitted by power laws, i.e., $\sigma_{T_c}/\Delta T=A Ra^{\beta}$, despite the $Ra$ range being limited for the oscillation state and the chaotic state. The fitted $\beta$ and $A$ are listed in table \ref{tab:fit}. We denote the transitional $Ra$ for different flow states as $Ra_o$ for the transition from convection to the oscillation state, $Ra_{ch}$ for the transition from the oscillation state to the chaotic state and $Ra_t$ for the transition from the chaotic state to the turbulent state. The so-determined transitional $Ra$ are $Ra_o=4.21\times10^4$, $Ra_{ch}=6.82\times10^4$, and $Ra_t=1.05\times10^5$. The $Ra$ range of different flow states is also summarised in table \ref{tab:fit}. Compared to the transitional values of $Ra$ between different flow states in nitrogen gas reported by \cite{Wei2021}, these values in liquid metal convection are systematically smaller. Note that $\sigma_{T_c}/\Delta T$ measured in cell B is systematically larger than that in cell A when there is an overlap of $Ra$. The reason for this difference remains unknown to us at present. For ease of data analysis, the data in cell B are multiplied by 0.8 to make them overlap with the data in cell A. The $\sigma_{T_c}/\Delta T$ scaling changes around $Ra=1\times10^6$ in the turbulent state, we refer to \S\ref{HT} for discussion on this transition.

\begin{table}
  \begin{center}
\def~{\hphantom{0}}
  \begin{tabular}{lccc}
       Flow state        & $Ra$ range                                        &   $A$                         & $\beta$\\[3pt]
       Conduction      & $Ra<5208$                                        &   -                              &  -     \\
       Convection      & $5208\leq4.21\times10^4$                        &   -                              &  -     \\
       Oscillation       & $4.21\times10^4\leq Ra<6.82\times10^4$    & 5.56                          & -0.80\\
       Chaotic           & $6.82\times10^4\leq Ra<1.05\times10^5$  & $5.85\times 10^{-32}$& 5.82\\
       Transition to turbulence     & $1.05\times10^5\leq Ra<10^6$  & $5.75\times 10^{-6}$  & 0.64\\
       Fully-developed turbulence     & $Ra\geq10^6$                           & 0.078                        & -0.05\\
       \end{tabular}
  \caption{Scaling of the temperature fluctuation $\sigma_{T_c}/\Delta T=A Ra^{\beta}$ for different flow states.}
  \label{tab:fit}
  \end{center}
\end{table}

Due to the limitation of cell size, we can not reach $Ra$ smaller than $1.17\times 10^4$. Thus the onset Rayleigh number for convection $Ra_c$ can not be determined directly. Following \cite{Wei2021}, we determine $Ra_c$ from the normalised time-averaged flow strength $\langle \delta\rangle /\Delta T$. When convection sets in, the first unstable mode is the azimuthal $m=1$ mode in a cylindrical cell with $\Gamma=1$ \citep{Hebert2010, Ahlers2022}. Thus $Ra_c$ can be determined once $\langle \delta\rangle /\Delta T$ is larger than the experimentally detectable temperature differences. The so-obtained $Ra_c$ is $Ra_c=5208$ (figure \ref{fig2}(b)). \cite{Ahlers2022} predicted that the critical Rayleigh number for the onset of convection is $Ra_{c,\Gamma} \sim 1708(1+0.77/{\Gamma}^{2})^2$. Substitute $\Gamma=1$ into the above relation, we obtain $Ra_c=5350$, which is close to the experimentally determined value. Hence, the  transitional values of $Ra$ between different flow states can be expressed as $Ra_o=8.08Ra_c$, $Ra_{ch}=13.10Ra_c$ and $Ra_t=20.16Ra_c$. Note \cite{Buell1983} reported that $Ra_c$ for a cylindrical cell with a perfectly adiabatic wall is $Ra_c \sim 3800$ theoretically, which is  37\% less than the experimentally determined value. 

\cite{Verzicco1997} studied the flow state evolution in mercury ($Pr=0.022$) in $\Gamma=1$ cylindrical cell. Similar flow evolutions are observed. But $Ra_c$, $Ra_o$, $Ra_{ch}$ and $Ra_t$ determined from DNS are all smaller than the values from the present study. The difference may be due to the thermal BCs on the sidewall or top/bottom plate. In the DNS study, the sidewall is adiabatic and the top/bottom plate is isothermal, but this is not the case for experimental studies, in which the sidewall is approximately adiabatic and the top/bottom plate is constant temperature/heat flux, nominally. The slight difference in $Pr$ between mercury and GaInSn may also play a role.

\subsection{Origin of the large-scale circulation (LSC) in the turbulent state}\label{subsec2}

\begin{figure}
  \centerline{\includegraphics[width=0.8\textwidth]{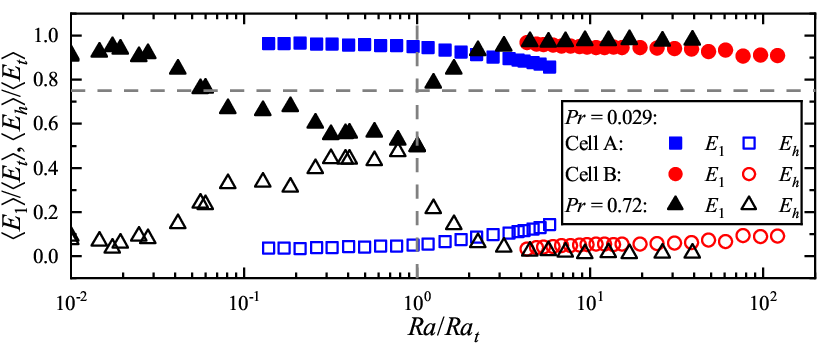}}
  \caption{Time-averaged energy ratio of the first Fourier mode to all Fourier modes $\langle E_1\rangle/\langle E_t \rangle$ and that of the high-order Fourier modes to all Fourier modes $\langle E_h\rangle/\langle E_t \rangle$ as a function of $Ra/Ra_t$. Here $\langle E_h\rangle=\sum_{n=2}^{4}\langle E_n \rangle$ and $\langle E_t\rangle=\sum_{n=1}^{4}\langle E_n \rangle$. Squares and circles are data measured in liquid metal with $Pr=0.029$ from the present study and triangles are data measured in nitrogen gas with $Pr=0.72$ from \cite{Wei2021}. $Ra_t$ is the critical Rayleigh number when the system enters the turbulent state. }
\label{fig3}
\end{figure}

One important issue in the study of turbulent convection is the origin of the LSC. \cite{Xi2004} showed that the self-organisation of thermal plumes is responsible for the initiation of LSC for a working fluid with $Pr\geq1$. However, the origin of the LSC in liquid metal convection remains unexplored. Figure \ref{fig2}(a) shows that similar to the observation in fluids with $Pr=0.72$ \citep{Wei2021}, the flow strength of the first Fourier mode increases to its maximum value just before the transition to turbulence and starts to decrease in the turbulent state. By analysing the energy of the different Fourier modes, we will show that the LSC is a residual of the cell structure observed near the onset of convection in liquid metal.

\begin{figure}
  \centerline{\includegraphics[width=\textwidth]{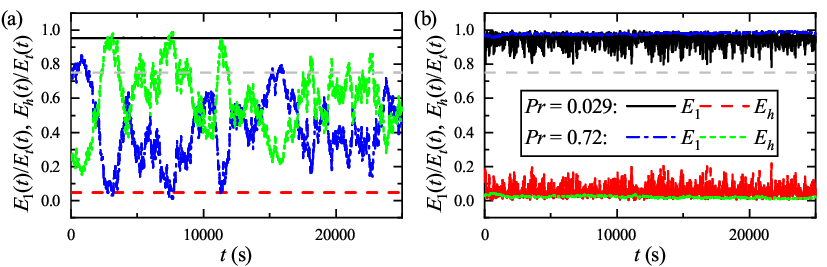}}
  \caption{Time series of $E_1(t)/E_t(t)$ and $E_h(t)/E_t(t)$ at (a) $Ra/Ra_t=0.76$, and (b) $Ra/Ra_t=13.1$. Legends of (a,b) are the same. Data for $Pr=0.72$ are adopted from \cite{Wei2021}.}
\label{fig4}
\end{figure}

Figure \ref{fig3} shows the time-averaged energy ratio of the first Fourier mode to all Fourier modes $\langle E_1\rangle/\langle E_t \rangle$ and that of the high-order Fourier modes to all Fourier modes $\langle E_h\rangle/\langle E_t \rangle$ as a function of $Ra/Ra_t$ for $Pr=0.029$ (squares and circles) and $Pr=0.72$ (triangles). Here $\langle E_h\rangle=\sum_{n=2}^{4}\langle E_n \rangle$ and $\langle E_t\rangle=\sum_{n=1}^{4}\langle E_n \rangle$. Measurements at different $z$ show similar dynamics, we will thus only focus on $z=H/2$. \cite{Wei2021} suggests that when $\langle E_1\rangle/\langle E_t\rangle<0.75$, shown as horizontal dashed lines in figure \ref{fig3}, the $E_1$ mode is considered as unstable. We follow this criterion to study the flow modes. It is seen that $\langle E_1\rangle/\langle E_{t}\rangle$ is always larger than 0.85 in liquid metal convection, suggesting the LSF is in the form of a single-roll structure. When $Ra<Ra_t$, the LSF is the cell structure observed just beyond the onset of convection, i.e., the $m=1$ azimuthal mode \citep{Hebert2010, Ahlers2022}. When $Ra>Ra_t$, the LSF is termed as the LSC. $\langle E_1\rangle/\langle E_{t}\rangle > 0.85$ for all $Ra$ suggests that the LSC in liquid metal convection is a residual of the cell structure near the onset of convection. To ascertain if the LSF will collapse over time, we checked the time trace of $E_1(t)/E_t(t)$ for $Ra<Ra_t$ measured in liquid metal convection and found that $E_1(t)/E_t(t)$ remains above 0.95. An example of the time trace of $E_1(t)/E_t(t)$ and $E_h(t)/E_t(t)$ for $Ra/Ra_t=0.76$ in liquid metal is shown in Figure \ref{fig4}($a$). It is seen that $E_1(t)/E_t(t)$ remains stable above 0.95 during a measurement period of $ \sim$ 7 hours, corresponding to 8150$\tau_{ff}$. 

The observation in liquid metal contrasts sharply with that in nitrogen gas with $Pr=0.72$. Figure \ref{fig3} shows that the energy of the cell structure $\langle E_1\rangle/\langle E_{t}\rangle$ in nitrogen gas decreases below 0.75 before the transition to turbulence. The energy of the LSC grows after the flow becomes turbulent. During the transition to turbulence $0.09\lesssim Ra/Ra_t \lesssim1$, the energy of the high-order modes is comparable with that of the first mode. This can be seen more clearly from the time series of $E_1(t)/E_t(t)$ and $E_h(t)/E_t(t)$ in nitrogen shown in figure \ref{fig4}(a). The data shows that the first mode and the high-order modes dominate the flow alternatively, suggesting that during the transition to turbulence, the cell structure is replaced by the high-order flow modes temporarily in fluids with $Pr\sim1$. The above results suggest that while the LSC in working fluids with $Pr\sim1$ originates from the self-organisation of thermal plume, the LSC in liquid metal originates from the cell structure. A possible reason for the different origins of the LSC may be the difference in thermal dissipation. Thermal bursts (plumes) will lose their heat content very quickly after they leave the thermal boundary layer in liquid metal, while they can keep their potential energy and convert it into kinetic energy while travelling upwards/downwards in working fluids with $Pr\sim1$, making the bursts strong enough to breakdown the cell structure. 

Figure \ref{fig3} shows that the time-averaged energy ratio of the LSC remains very close to 1 in turbulent RBC with both liquid metal and nitrogen as the working fluid. How about their time evolution? Figure \ref{fig4}(b) shows the time series of $E_1(t)/E_t(t)$ and $E_h(t)/E_t(t)$ at $Ra/Ra_t=13.1$ for $Pr=0.029$ and $Pr=0.72$. For both cases, $E_1(t)/E_t(t)$ is larger than 0.75, suggesting that the LSC is the dominant flow mode in the turbulent state. However, much more intensive fluctuation in $E_1(t)/E_t(t)$ is observed in liquid metal than that in nitrogen gas.

\subsection{The heat transport and temperature fluctuation at the cell centre}\label{HT} 

\begin{figure}
  \centerline{\includegraphics[width=\textwidth]{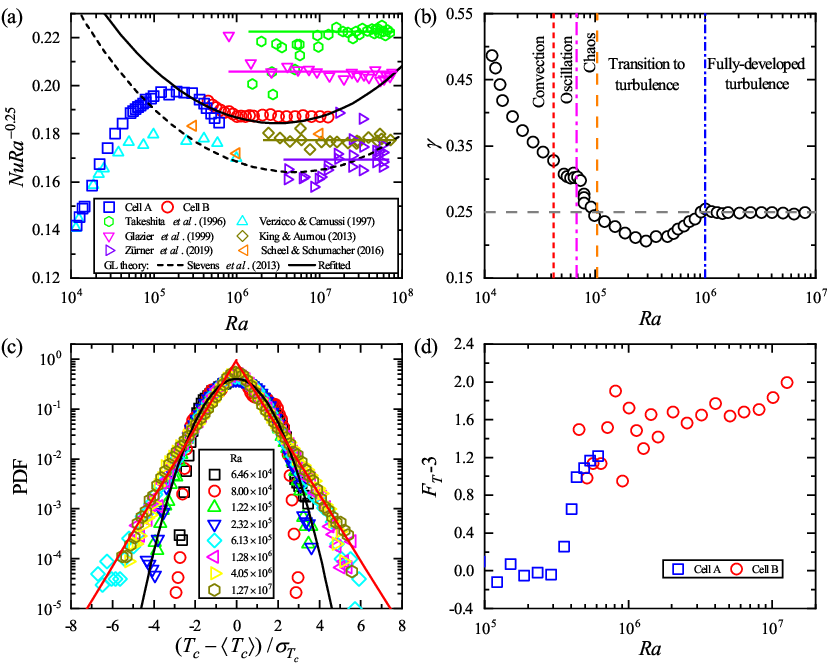}}
  \caption{(a) Compensated plot of $Nu$ vs $Ra$. (b) The effective heat transport scaling exponent $\gamma$ vs $Ra$. (c) PDF and (d) flatness $F_T-3$ of the temperature fluctuation at the cell centre.}
\label{fig5}
\end{figure}

The flow state evolution discussed in \S\ref{subsec1} is also reflected by the heat transport and the temperature fluctuation at the cell centre. Figure \ref{fig5}(a) plots $Nu/Ra^{0.25}$ vs $Ra$. It is seen from the present measurements (squares and circles) that the heat transport scaling changes when the flow evolves from convection to oscillation, chaos and turbulence. We determined local effective scaling exponents, i.e., $Nu=BRa^{\gamma}$, using a sliding window that covers about half a decade in $Ra$. Figure \ref{fig5}(b) shows $\gamma$ as a function of $Ra$. One sees that $\gamma$ continues to evolve when the flow state changes. Near the onset of convection at $Ra=10^4$, one obtains $\gamma=0.49$. While for $Ra>10^6$, one obtains $\gamma=0.25$, and $\gamma$ remains at 0.25 up to the maximum $Ra$ achieved. Concurrently occurring with the changing of the heat transport scaling is the shape of the probability density function (PDF) of the temperature fluctuation at the cell centre shown in figure \ref{fig5}(c). It is seen that the PDFs can be approximately by Gaussian functions for $1.22\times10^5<Ra<10^6$, while an exponential function fits the data better for $Ra>10^6$, beyond which the shapes of the PDF remain almost invariant with $Ra$, suggesting that the flow has entered a universal state. To quantify the shape change of the PDFs, we plot in figure \ref{fig5}(d) the flatness $F_T-3$ of $T_c$. Here $F_T$ is defined as $ F_{T}= {\left \langle {(T_c-\left \langle T_c \right \rangle)}^{4}\right \rangle}/\sigma_{T_c}^{4}$. It is seen that during the transition to turbulence, $F_T-3$ is close to 0, consistent with the PDFs being of Gaussian shape. $F_T$ increases with $Ra$ until $Ra>10^6$. When the flow becomes fully turbulent, $F_T$ remains almost at a constant around $F_T=4.7$. The observation of universal form of the PDFs for $Ra>10^6$ is consistent with $\gamma=0.25$ for $Ra>10^6$. This flow state transition in the turbulent state may also be responsible for the change of temperature fluctuation scaling shown in figure \ref{fig2}.

We now compare our measurements with published data in the literature. The dashed line in figure \ref{fig5}(a) represents the prediction of the Grossman-Lohse (GL) theory \citep{Grossmann2000}, using the updated pre-factors from \cite{GL2013}, which describes $Nu(Ra, Pr)$ and $Re(Ra, Pr)$ with the two coupled equations (\ref{Equ1}) and (\ref{Equ2}),

\begin{align}
(Nu-1)RaPr^{-2} & = c_1 \frac{Re^{2}}{g(\sqrt{{Re_L}/{Re}})}+c_2 Re^{3}, \label{Equ1}\\
Nu-1 & =  c_3 Pr^{1/2}Re^{1/2}\left\{f\left[\frac{2aNu}{\sqrt{Re_L}}g\left(\sqrt{\frac{Re_L}{Re}}\right)\right]\right\}^{1/2}\notag \\
&+c_4 PrRef\left[\frac{2aNu}{\sqrt{Re_L}}g\left(\sqrt{\frac{Re_L}{Re}}\right)\right],\label{Equ2}
\end{align}

where the $f$ and $g$ are the cross-over functions, $c_1=8.05$, $c_2=1.38$, $c_3=0.487$, $c_4=0.0252$, $a=0.922$, and $Re_L=(2a)^2$. The data follow well the GL theory's general trend in the turbulent state. But the predicted $Nu$ is systematically lower than the measured $Nu$. So we refit the GL parameters using $Nu$ measured at $Ra=1.02 \times 10^7$ and $Pr=0.029$ from the present study instead of the data point at $Ra=1\times10^7$ and $Pr=0.025$ from \cite{Cioni1997}. The so-obtained fitting parameters are $c_1=18.60$, $c_2=3.62$, $c_3=0.62$, and $c_4=0.04$. The refitted GL theory prediction is plotted as a black solid line in figure \ref{fig5}(a). It is seen that the refitted GL theory is in good agreement with measurements in the entire turbulent state.

Figure \ref{fig5}(a) also plots $Nu$ data from experiments and DNS in liquid metal convection. To exclude the effects of $\Gamma$ and cell geometry on $Nu$, we only consider measurements performed in cylinders with $\Gamma=1$. It is seen that before the transition to turbulence ($Ra<1\times 10^5$),  present measurements are in good agreement with the DNS data by \cite{Verzicco1997}. In the fully-developed turbulent state, on one hand, our data is larger by $\sim 10\%$ than the DNS data from \cite{Verzicco1997} and \cite{Scheel2016}. On the other hand, our data can be regarded as consistent with other experimental data, i.e., to the lowest order, all data exhibits a $Nu\sim Ra^{0.25}$ scaling for $Ra>1\times10^6$ as indicated by the horizontal lines. The pre-factor $B$ from different experimental studies varies from 0.17 to 0.22, which may be due to differences in the temperature BCs on the sidewall or those on the top plate and the bottom plate. As for the temperature BCs on the plates from the present study and studies reported by \citep{Takeshita1996, Glazier1999, King2013, Zurner2019}, the top plate is isothermal and the bottom plate is constant heat flux, nominally. However, in the present study, the sidewall BC using the Plexiglas (the wall admittance $C=1038, 2248$ for cell A and cell B, respectively) is much more close to the adiabatic condition compared with the other experiments using stainless steel ($C<67$) \citep{Takeshita1996, Glazier1999, King2013}.  One should also notice that $Nu$ variation may be caused by the different thermal insulation methods used in different studies.

\section {Conclusion and outlook}\label{sec:con}

We have studied the flow state and heat transport in liquid metal Rayleigh-B\'enard convection. The experiments were performed in two cylindrical cells with $\Gamma=1$. Using liquid gallium-indium-tin alloy as the working fluid, the experiments covered a Rayleigh number range from $1.17\times10^4$ to $1.27\times 10^7$ and at a fixed Prandtl number of 0.029. The temperature at the cell centre was used as a proxy for the flow state. It is found that with increasing $Ra$, the flow evolves from the convection state to the oscillation state, the chaotic state and finally to a turbulent state for $Ra>10^5$. The transition to turbulence is found to follow a period-doubling bifurcation scenario. The observed flow evolution confirms the DNS results by \cite{Verzicco1997}. In addition, it is found that the LSC in the turbulent state is a residual of the cell structure in liquid metal convection. This observation contrasts sharply with the results obtained in turbulent convection with $Pr\sim1$, where the cell structure near the onset of convection was temporarily replaced by high-order flow modes before the transition to turbulence. The evolution of the flow state is also reflected by the Nusselt number $Nu$ and the PDF of the temperature fluctuation at the cell centre. It is found that the effective heat transport scaling exponent $\gamma$, i.e., $Nu\sim Ra^{\gamma}$, changes continuously from $\gamma=0.49$ at $Ra\sim 10^4$ to $\gamma=0.25$ for $Ra>10^6$. Meanwhile, the PDF at the cell centre gradually evolves from a Gaussian-like shape to an exponential-like shape in the fully-developed turbulent state. For $Ra>10^6$, the flow shows self-similar behaviour, which is revealed by the universal shape of the PDF of the temperature fluctuation at the cell centre and $Nu=0.19Ra^{0.25}$ for heat transport. Finally, we remark that the findings presented are obtained in cells with $\Gamma=1$. However, $\Gamma$ plays a vital role in determining the flow dynamics and heat transport in liquid metal convection \citep{Schindler2022}. Thus it will be worthwhile studying the effects of $\Gamma$ on the flow evolution and heat transport.

\backsection[Acknowledgements]{We thank P. Wei for making the data in figure \ref{fig3} and figure \ref{fig4} available to us. We are also grateful to her, Y. Wang, and G.-Y. Ding for discussion.}

\backsection[Funding]{This work was supported by the National Natural Science Foundation of China (NSFC) (Grant Nos 12002260 and 92152104) and the Fundamental Research Funds for the Central Universities. }
 
\backsection[Declaration of interests] {The authors report no conflict of interest.}

\backsection[Author ORCID]{\\Lei Ren, https://orcid.org/0000-0001-8256-0834;\\Xin Tao, https://orcid.org/0000-0002-7070-0788;\\Lu Zhang, https://orcid.org/0000-0003-4009-2969;\\Ming-Jiu Ni, https://orcid.org/0000-0003-3699-8370;\\ Ke-Qing Xia, https://orcid.org/0000-0001-5093-9014;\\Yi-Chao Xie, https://orcid.org/0000-0002-2159-4579.}

\bibliographystyle{jfm}
\bibliography{jfm}

\end{document}